\documentclass[conference,a4paper]{IEEEtran}

\normalsize

\ifCLASSINFOpdf
\else
\fi

\usepackage{amsmath}
\usepackage{amssymb}
\usepackage{cite}
\usepackage{graphicx}
\usepackage{subfigure}
\usepackage{multirow} 
\usepackage{longtable}
\usepackage{threeparttable}
\usepackage{caption3}
\usepackage{array}
\usepackage{cases}
\usepackage{bbm}
\usepackage{color}
\usepackage[T1]{fontenc}
\usepackage[utf8]{inputenc}
\usepackage{authblk}

\newcommand\mycom[2]{\genfrac{}{}{0pt}{}{#1}{#2}}

\usepackage[absolute]{textpos} 
\usepackage{blindtext}

\begin{document}


\title{Waveform Optimization for Wireless Power Transfer with Nonlinear Energy Harvester Modeling}

\author[1,3]{Bruno Clerckx}
\author[1]{Ekaterina Bayguzina}
\author[2]{David Yates}
\author[2]{Paul D. Mitcheson}
\affil[1]{Comm. and Sig. Proc. Group, EEE Department, Imperial College London, United Kingdom}
\affil[2]{Control and Power Group, EEE Department, Imperial College London, United Kingdom}
\affil[3]{School of Electrical Engineering, Korea University, Korea}
\affil[ ]{Email:\{b.clerckx,ekaterina.bayguzina08,david.yates,paul.mitcheson\}@imperial.ac.uk}

\maketitle

\begin{abstract} Far-field Wireless Power Transfer (WPT) and Simultaneous Wireless Information and Power Transfer (SWIPT) have attracted significant attention in the RF and communication communities. Despite the rapid progress, the problem of waveform design to enhance the output DC power of wireless energy harvester has received limited attention so far. In this paper, we bridge communication and RF design and derive novel multisine waveforms for multi-antenna wireless power transfer. The waveforms are adaptive to the channel state information and result from a posynomial maximization problem that originates from the non-linearity of the energy harvester. They are shown through realistic simulations to provide significant gains (in terms of harvested DC power) over state-of-the-art waveforms under a fixed transmit power constraint. 
\footnote{This work has been partially supported by the EPSRC of the UK under grant EP/M008193/1, EP/K502856/1 and EP/L504786/1.}
\end{abstract}


\IEEEpeerreviewmaketitle

\section{Introduction}


\IEEEPARstart{W}{ireless} Power Transfer (WPT) via radio-frequency radiation is attracting more and more attention. RF radiation has indeed become a viable source for energy harvesting with clear applications in Wireless Sensor Networks (WSN) and Internet of Things (IoT) \cite{Visser:2013,Pinuela:2013}. Interestingly, since WPT and wireless communication share the same RF medium to transfer power and information, the emerging field of Simultaneous Wireless Information and Power Transfer (SWIPT) has recently attracted significant attention in academia \cite{Zhang:2013,Park:2013}. 
\par The challenge with WPT is to find ways to increase the DC power level at the output of the energy harvester without increasing the transmit power. Most of the technical efforts have been devoted to the design of efficient energy harvesters (so-called rectenna) \cite{Visser:2013,Pinuela:2013}. Interestingly, the overall RF-to-DC conversion efficiency of the rectenna is not only a function of its design but also of its input waveform. However, the waveform design has received less attention \cite{Trotter:2009,Boaventura:2011,Collado:2014}. In \cite{Trotter:2009,Boaventura:2011}, a multisine signal excitation is shown through analysis, simulations and measurements to enhance the DC power and RF-DC conversion efficiency over a single sinewave signal. In \cite{Collado:2014}, various input waveforms (OFDM, white noise, chaotic) are considered and experiments show that waveforms with high peak to average power ratio (PAPR) increase RF-to-DC conversion efficiency. Even though those papers provide some useful insights into the impact of waveform design onto WPT performance, there are many limitations in the WPT waveform design literature: 1) there has not been any formal tool to optimize waveforms for WPT so far, 2) multipath fading (well known in wireless communications) has been ignored despite its tremendous impact on the received waveform at the input of the energy harvester, 3) the Channel State Information (CSI) is assumed unknown to the transmitter, 4) the transmitter is commonly equipped with a single antenna. 
\par In the rapidly expanding SWIPT community, the non-linearity of the energy harvester is not accurately modeled. It is indeed assumed that the harvested DC power is modeled as a conversion efficiency constant multiplied by the average power of the input signal to the energy harvester \cite{Zhang:2013,Park:2013}. This was used so as to simplify the design of SWIPT but is unfortunately unrealistic \cite{Trotter:2009,Boaventura:2011,Collado:2014}. 
\par In this paper we tackle the important problem of waveform optimization for Multiple Input-Single Output (MISO) WPT. We consider WPT with a multisine waveform (due to its popularity in communication, e.g.\ OFDM) on each antenna and transmission over a multipath channel. We introduce a simple and tractable analytical model of the energy havester, accounting for its nonlinearity. Assuming perfect Channel State Information at the Transmitter (CSIT) can be attained (in a similar way as it is done in wireless communication systems), we formulate an optimization problem to adaptively change on each transmit antenna the multisine waveform as a function of the CSI so as to maximize the output DC current at the energy harvester. The globally optimal phases of multisine waveform weights are obtained in closed form while the locally optimal amplitudes are shown to result from a non-convex posynomial maximization problem subject to a power constraint, which can be formulated as a Reverse Geometric Programming and solved iteratively. The optimized waveforms adaptive to the CSI are shown to provide significant gains over state-of-the-art waveforms.
\par \textit{Organization:} Section \ref{system_model} introduces the WPT system model and section \ref{section_EH} the analytical model of the energy harvester. Section \ref{section_waveform} tackles the waveform optimization and section \ref{simulations} evaluates the performance. Section \ref{conclusions} concludes the work.
\par \textit{Notations:} Bold lower case and upper case letters stand for vectors and matrices respectively whereas a symbol not in bold font represents a scalar. Operator $\left\|.\right\|_F^2$ refers to the Frobenius norm a matrix. $\mathcal{E}\left\{.\right\}$ refers to the expectation operator.

\section{WPT System Model}\label{system_model}
Consider the transmitter made of $M$ antennas and $N$ sinewaves whose multisine transmit signal at time $t$ on transmit antenna $m=1,\ldots,M$ is given by
\begin{equation}\label{WPT_waveform}
x_m(t)=\sum_{n=0}^{N-1} s_{n,m} \cos(w_n t+\phi_{n,m})
\end{equation}
where $s_{n,m}$ and $\phi_{n,m}$ refer to the amplitude and phase of the $n^{th}$ sinewave at frequency $w_n$ on transmit antenna $m$, respectively. We assume for simplicity that the frequencies are evenly spaced, i.e.\ $w_n=w_0+n\Delta_w$ with $\Delta_w$ the frequency spacing. The magnitudes and phases of the sinewaves can be collected into matrices $\mathbf{S}$ and $\mathbf{\Phi}$ such that the $(n,m)$ entries of $\mathbf{S}$ and $\mathbf{\Phi}$ write as $s_{n,m}$ and $\phi_{n,m}$, respectively. The transmitter is subject to the power constraint $\sum_{m=1}^{M}\mathcal{E}\big\{\left|x_m\right|^2\big\}=\frac{1}{2}\left\|\mathbf{S}\right\|_F^2\leq P$. 

\par The multi-antenna transmitted sinewaves propagate through a multipath channel, characterized by $L$ paths whose delay, amplitude, phase and direction of departure (chosen with respect to the array axis) are respectively denoted as $\tau_l$, $\alpha_l$, $\xi_l$ and $\theta_l$, $l=1,\ldots,L$. We assume transmit antennas are closely located so that $\tau_l$, $\alpha_l$ and $\xi_l$ are the same for all transmit antennas (assumption of a narrowband balanced array) \cite{Clerckx:2013}. The signal transmitted by antenna $m$ and received at the single-antenna receiver after multipath propagation can be written as
\begin{align}\label{received_signal_ant_m}
&y^{(m)}(t)\\
&=\sum_{n=0}^{N-1} s_{n,m}\left(\sum_{l=0}^{L-1}\alpha_l \cos(w_n(t-\tau_l)+\xi_l+\phi_{n,m}+\Delta_{n,m,l})\right)\nonumber
\end{align}
where $\Delta_{n,m,l}$ refers to the phase shift between the $m^{th}$ transmit antenna and the first one. For simplicity, we assume that $\Delta_{n,1,l}=0$. For a Uniform Linear Array (ULA), $\Delta_{n,m,l}=2\pi (m-1)\frac{d}{\lambda_n}\cos(\theta_l)$ where $d$ is the inter-element spacing, $\lambda_n$ the wavelength of the $n^{th}$ sinewave.

\par The quantity between the brackets in \eqref{received_signal_ant_m} can simply be rewritten as 
\begin{multline}
\sum_{l=0}^{L-1}\alpha_l \cos(w_n(t-\tau_l)+\xi_l+\phi_{n,m}+\Delta_{n,m,l})\\
=A_{n,m} \cos(w_n t+\psi_{n,m})
\end{multline}
where the amplitude $A_{n,m}$ and the phase $\psi_{n,m}$ are such that
\begin{align}\label{A_nm_psi}
A_{n,m}e^{j \psi_{n,m}}&=A_{n,m}e^{j \left(\phi_{n,m}+\bar{\psi}_{n,m}\right)}=e^{j \phi_{n,m}}h_{n,m}
\end{align}
with $h_{n,m}=A_{n,m}e^{j \bar{\psi}_{n,m}}=\sum_{l=0}^{L-1}\alpha_l e^{j(-w_n\tau_l+\Delta_{n,m,l}+\xi_l)}$ the frequency response of the channel of antenna $m$ at $w_n$. 
\par The total received signal comprises the sum over all transmit antennas, namely 
\begin{align}
y(t)=\sum_{m=1}^{M} y^{(m)}(t)&=\sum_{m=1}^{M}\sum_{n=0}^{N-1}s_{n,m}A_{n,m} \cos(w_n t+\psi_{n,m})\nonumber\\
&=\sum_{n=0}^{N-1}X_n \cos(w_n t+\delta_{n})\label{y_t}
\end{align}
where $X_n e^{j \delta_{n}}=\sum_{m=1}^M s_{n,m}A_{n,m}e^{j \psi_{n,m}}$.

\section{Analytical Model of the Energy Harvester and DC Component}\label{section_EH}
In this section we derive a simple and tractable model of the energy harvester circuit and express the output DC current as a function of the waveform parameters. The model relies on several assumptions that are made to make the model tractable and therefore be able to optimize the waveforms. Performance evaluations will be conducted in Section \ref{simulations} using a more accurate circuit simulator.

\subsection{Antenna Equivalent Circuit}
\begin{figure}
\centerline{\includegraphics[width=0.6\columnwidth]{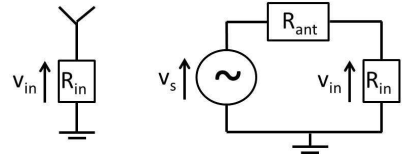}}
  \caption{Antenna equivalent circuit.}
  \label{antenna_model}
\end{figure}

Assume an energy harvester whose input impedance $R_{in}$ is connected to a receiving antenna as in Fig.\ \ref{antenna_model}. The signal $y(t)$ impinging on the antenna has an average power $P_{av}=\mathcal{E}\big\{y(t)^2\big\}$. Following \cite{Curty:2005}, the antenna is assumed lossless and modeled as an equivalent voltage source $v_s(t)$ in series with an impedance $R_{ant}=50\Omega$, as illustrated in Fig.\ \ref{antenna_model}. In the perfect matching case ($R_{in}=R_{ant}$), the received power $P_{av}$ is completely transferred to the energy harvester's input impedance such that $P_{av}=\mathcal{E}\big\{v_{in}(t)^2\big\}/R_{in}$ where $v_{in}(t)$ is the input voltage to the energy harvester. Under perfect matching, $v_{in}(t)$ is half of $v_s(t)$ and both can be related to the received signal $y(t)$ as
\begin{equation}
v_s(t)=2 y(t) \sqrt{R_{ant}}, \hspace{0.5cm} v_{in}(t)=y(t) \sqrt{R_{ant}}.
\end{equation} 

\subsection{Rectifier and Diode Non-Linearity}
\begin{figure}
\centerline{\includegraphics[width=0.6\columnwidth]{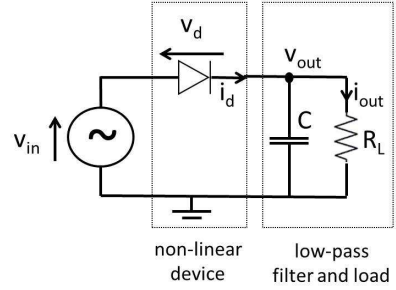}}
  \caption{A single diode rectifier.}
  \label{rectifier}
\end{figure}

A rectifier is always made of a non-linear device (e.g.\ diode) followed by a low pass filter (LPF) with load \cite{Pinuela:2013,Trotter:2009,Boaventura:2011}. A simplified rectifier circuit is illustrated in Fig.\ \ref{rectifier}. We assume that its input impedance has been perfectly matched to the antenna impedance. 

The current $i_d(t)$ flowing through an ideal diode (neglecting its series resistance) relates to the voltage drop across the diode $v_d(t)=v_{in}(t)-v_{out}(t)$ as $i_d(t)=i_s\big(e^{\frac{v_d(t)}{n v_t}}-1\big)$
where $i_s$ is the reverse bias saturation current, $v_t$ is the thermal voltage, $n$ is the ideality factor (assumed equal to 1 for simplicity). In order to express the non-linearity of the diode, we take a Taylor expansion of the exponential function around a fixed operating voltage drop $v_d=a$ such that the diode current can be equivalently written as
\begin{equation}\label{taylor}
i_d(t)=\sum_{i=0}^\infty k_i (v_d(t)-a)^i=\sum_{i=0}^\infty k_i (v_{in}(t)-v_{out}(t)-a)^i,
\end{equation}
where $k_0=i_s\big(e^{\frac{a}{n v_t}}-1\big)$ and $k_i=i_s\frac{e^{\frac{a}{n v_t}}}{i!\left(n v_t\right)^i}$, $i=1,\ldots,\infty$. 
As such, it is not easy to infer from \eqref{taylor} the exact dependencies of the diode current on the waveform parameters since both $v_{in}(t)$ and $v_{out}(t)$ will depend and fluctuate over time as a function of the waveform. Nevertheless, assuming a steady-state response, an ideal rectifier would deliver a constant output voltage $v_{out}$ that would track the largest peaks of the input voltage $v_{in}(t)$ \cite{Curty:2005}. As a consequence, the output current delivered to the load $i_{out}$ would also be constant. Denoting the magnitude of the peaks of $v_{in}(t)$ as $\hat{v}_{in}$, $v_{out}=\hat{v}_{in}$. In this ideal rectifier, since $v_{out}$ is a constant (we drop the time dependency), a suitable choice of the operating voltage drop $a$ would be $a=\mathcal{E}\left\{v_{in}(t)-v_{out}\right\}=-v_{out}=-\hat{v}_{in}$ since $\mathcal{E}\left\{v_{in}(t)\right\}=\sqrt{R_{ant}}\mathcal{E}\left\{y(t)\right\}=0$. Under such assumptions, \eqref{taylor} can simply be written as $i_d(t)=\sum_{i=0}^\infty k_i R_{ant}^{i/2} y(t)^i$, which now makes the dependency between the diode current $i_d(t)$, the received waveform $y(t)$ and therefore the transmitted waveforms $\left\{x_m(t)\right\}$ much more explicit. Moreover, with such a choice of $a=-v_{out}$, the Taylor expansion of the exponential function could be truncated without compromising too much on the accuracy. 

\subsection{Output DC Current}
The problem at hand will be the design of $\left\{x_m(t)\right\}$ such that the output DC current is maximized. Under the ideal rectifier assumption, the current delivered to the load in a steady-state response is constant and given by $i_{out}=\mathcal{E}\left\{i_d(t)\right\}$. 
\par In order to make the optimization tractable but keep the fundamental non-linear behaviour of the diode, we truncate the Taylor expansion to the fourth order, as argued in \cite{Boaventura:2011} to be sufficient to demonstrate the rectification operation. The output current therefore approximates as 
\begin{equation}\label{diode_model}
i_{out}=\mathcal{E}\left\{i_d(t)\right\}\approx \sum_{i=0}^4 k_i R_{ant}^{i/2} \mathcal{E}\left\{y(t)^i\right\}.
\end{equation}
Applying \eqref{y_t} to \eqref{diode_model} and taking the expectation over time, we get an approximation of the DC component of the current at the output of the rectifier (and the low-pass filter) with a multisine excitation over a multipath channel as $i_{out}=k_0+z_{DC}$ where 
\begin{equation}\label{diode_model_2}
z_{DC}=k_2 R_{ant} \mathcal{E}\left\{y(t)^2\right\}+k_4 R_{ant}^2 \mathcal{E}\left\{y(t)^4\right\}
\end{equation}
is detailed in \eqref{y_DC} (at the top of next page). There is no third order term since $\mathcal{E}\left\{y(t)^3\right\}=0$. Expression \eqref{diode_model_2} was used in \cite{Boaventura:2011} for single-antenna WPT to show the suitability of in-phase multisine excitation with uniform power allocation for scenarios without multipath and in the absence of CSIT.

\begin{table*}
\begin{align}\label{y_DC}
z_{DC}(\mathbf{S},\mathbf{\Phi})&=\frac{k_{2}}{2}R_{ant}\left[\sum_{n=0}^{N-1} \sum_{m_0,m_1} s_{n,m_0}s_{n,m_1}A_{n,m_0}A_{n,m_1}\cos\left(\psi_{n,m_0}-\psi_{n,m_1}\right)\right]\nonumber\\
&\hspace{1cm}+\frac{3k_{4}}{8}R_{ant}^2\left[\sum_{\mycom{n_0,n_1,n_2,n_3}{n_0+n_1=n_2+n_3}}\sum_{\mycom{m_0,m_1,}{m_2,m_3}}\Bigg[\prod_{j=0}^3s_{n_j,m_j}A_{n_j,m_j}\Bigg]\cos(\psi_{n_0,m_0}+\psi_{n_1,m_1}-\psi_{n_2,m_2}-\psi_{n_3,m_3})\right].
\end{align}\hrulefill
\end{table*}

\par It is worth contrasting \eqref{diode_model_2} with the model commonly used in the SWIPT literature. In \cite{Zhang:2013,Park:2013}, the total harvested power is defined as $E=\zeta\mathcal{E}\left\{y(t)^2\right\}$ where $\zeta$ is a constant that refers to the RF-to-DC conversion efficiency and is commonly chosen as 1 for simplicity. Comparing with \eqref{diode_model_2}, we note that $E$ only accounts for the second order term in the Taylor expansion and ignores the non-linearity of the rectifier (modeled by the fourth order term). The maximization of $E$ subject to the transmit power constraint would lead to a single-sinewave transmission strategy where the power is allocated to the strongest sinewave, i.e.\ the one corresponding to the strongest channel \cite{Bayani:2015}. This contrasts with RF experiments that highlight the usefulness of allocating power to multiple sinewaves \cite{Trotter:2009,Boaventura:2011,Collado:2014}. If non-linearity is accounted for, the presence of the fourth order term in \eqref{diode_model_2} suggests that transmitting over a single sinewave is in general suboptimal.

\section{Waveform Optimization}\label{section_waveform}
Assuming the rectifier characteristics $k_0,k_2,k_4$ and the channel state information (in the form of frequency response $h_{n,m}$) are known to the transmitter, we now aim at finding the optimal set of amplitudes and phases (across antennas and frequencies) $\mathbf{S},\mathbf{\Phi}$ that maximize $i_{out}$, i.e.\
\begin{align}\label{P1}
\max_{\mathbf{S},\mathbf{\Phi}} \hspace{0.3cm} &i_{out}(\mathbf{S},\mathbf{\Phi})\\
\textnormal{subject to} \hspace{0.3cm} &\frac{1}{2}\left\|\mathbf{S}\right\|_F^2\leq P.
\end{align}
From the previous section, we however note that $k_0,k_2,k_4$ are functions of $a=-v_{out}$ which is affected by the choice of $\mathbf{S},\mathbf{\Phi}$. Similarly, $\mathbf{S},\mathbf{\Phi}$ are functions of $k_0,k_2,k_4$. This suggests that $k_0,k_2,k_4$ and $\mathbf{S},\mathbf{\Phi}$ should be iteratively computed by fixing $k_0,k_2,k_4$ when $\mathbf{S},\mathbf{\Phi}$ are optimized and inversely. We here assume that $k_0,k_2,k_4$ have been computed for a given $a$ and we aim at finding the optimal $\mathbf{S},\mathbf{\Phi}$. For a fixed $a$, $k_0$ is fixed and is not affected by the choice of $\mathbf{S},\mathbf{\Phi}$. Hence the problem \eqref{P1} can equivalently be re-written as
\begin{align}\label{P2}
\max_{\mathbf{S},\mathbf{\Phi}} \hspace{0.3cm} &z_{DC}(\mathbf{S},\mathbf{\Phi})\\
\textnormal{subject to} \hspace{0.3cm} &\frac{1}{2}\left\|\mathbf{S}\right\|_F^2\leq P.
\end{align}
Interestingly, the optimal phases can be obtained first in closed form. Given the optimal phases, the optimal amplitudes can be computed numerically.

\subsection{Phase Optimization}

To maximize $z_{DC}(\mathbf{S},\mathbf{\Phi})$, we should aim at guaranteeing that all $\cos(.)$ are equal to 1 in \eqref{y_DC}. This can be satisfied by choosing $\psi_{n,m}=0$ $\forall n,m$ (and therefore $\delta_{n}=0$ $\forall n$), which implies from \eqref{A_nm_psi} to choose the optimal sinewave phases as $\phi_{n,m}^{\star}=-\bar{\psi}_{n,m}$.
$\mathbf{\Phi}^{\star}$ is obtained by collecting $\phi_{n,m}^{\star}$ $\forall n,m$ into matrix.

\subsection{Amplitude Optimization}
With the optimal phases, $\psi_{n,m}=0$ and $X_n=\sum_{m=1}^M s_{n,m}A_{n,m}$ such that $z_{DC}(\mathbf{S},\mathbf{\Phi}^{\star})$ is simply obtained from \eqref{y_DC} by replacing all $\cos(.)$ functions by 1.

\par Recall from \cite{Chiang:2005} that a monomial is defined as the function $g:\mathbb{R}_{++}^{N}\rightarrow\mathbb{R}:g(\mathbf{x})=c x_1^{a_1}x_2^{a_2}\ldots x_N^{a_N}$
where $c>0$ and $a_i\in\mathbb{R}$. A sum of $K$ monomials is called a posynomial and can be written as $f(\mathbf{x})=\sum_{k=1}^K g_k(\mathbf{x})$ with $g_k(\mathbf{x})=c_k x_1^{a_{1k}}x_2^{a_{2k}}\ldots x_N^{a_{Nk}}$
where $c_k>0$. As we can see from \eqref{y_DC}, $z_{DC}(\mathbf{S},\mathbf{\Phi}^{\star})$ is a posynomial.

The optimization problem becomes
\begin{align}
\max_{\mathbf{S}} \hspace{0.3cm}&z_{DC}(\mathbf{S},\mathbf{\Phi}^{\star})\\
\textnormal{subject to} \hspace{0.3cm} &\frac{1}{2}\left\|\mathbf{S}\right\|_F^2\leq P.
\end{align}
It therefore consists in maximizing a posynomial subject to a power constraint (which itself is written as a posynomial). 
Unfortunately this problem is not a standard Geometric Program (GP) but it can be transformed to an equivalent problem by introducing an auxiliary variable $t_0$
\begin{align}\label{reverse_GP}
\min_{\mathbf{S},t_0} \hspace{0.3cm} &1/t_0\\
\textnormal{subject to} \hspace{0.3cm} &\frac{1}{2}\left\|\mathbf{S}\right\|_F^2\leq P,\\ 
&z_{DC}(\mathbf{S},\mathbf{\Phi}^{\star})/t_0\geq 1.\label{reverse_GP_2}
\end{align}

This is known as a Reverse Geometric Program due to the minimization of a posynomial subject to upper and lower bounds inequality constraints \cite{Duffin:1973,Chiang:2005}. Note that  $z_{DC}(\mathbf{S},\mathbf{\Phi}^{\star})/t_0\geq 1$ is equivalent to $t_0/z_{DC}(\mathbf{S},\mathbf{\Phi}^{\star})\leq1$. However $1/z_{DC}(\mathbf{S},\mathbf{\Phi}^{\star})$ is not a posynomial, therefore preventing the use of standard GP tools.
The idea is to lower bound $z_{DC}(\mathbf{S},\mathbf{\Phi}^{\star})$ by a monomial $\bar{z}_{DC}(\mathbf{S})$, i.e.\ upper bound $1/z_{DC}(\mathbf{S},\mathbf{\Phi}^{\star})$ by the monomial $1/\bar{z}_{DC}(\mathbf{S})$ (since the inverse of a monomial is still a monomial) \cite{Duffin:1970,Duffin:1973}. Let $\left\{g_k(\mathbf{S},\mathbf{\Phi}^{\star})\right\}$ be the monomial terms in the posynomial $z_{DC}(\mathbf{S},\mathbf{\Phi}^{\star})=\sum_{k=1}^K g_k(\mathbf{S},\mathbf{\Phi}^{\star})$. The choice of the lower bound relies on the fact that an arithmetic mean is greater or equal to the geometric mean. Hence, $z_{DC}(\mathbf{S},\mathbf{\Phi}^{\star})\geq \prod_{k=1}^K\left(g_k(\mathbf{S},\mathbf{\Phi}^{\star})/\gamma_k\right)^{\gamma_k}=\bar{z}_{DC}(\mathbf{S})$,
where $\gamma_k\geq 0$ and $\sum_{k=1}^K \gamma_k=1$. 
Since
\begin{equation}\label{UB}
1/z_{DC}(\mathbf{S},\mathbf{\Phi}^{\star})\leq 1/\bar{z}_{DC}(\mathbf{S}),
\end{equation}
we can replace (in a conservative way) inequality $t_0/z_{DC}(\mathbf{S},\mathbf{\Phi}^{\star})\leq1$ by $t_0/\bar{z}_{DC}(\mathbf{S})=t_0\prod_{k=1}^K\left(g_k(\mathbf{S},\mathbf{\Phi}^{\star})/\gamma_k\right)^{-\gamma_k}\leq1$. For a given choice of $\left\{\gamma_k\right\}$, problem \eqref{reverse_GP}-\eqref{reverse_GP_2} is now replaced by the standard GP
\begin{align}\label{standard_GP}
\min_{\mathbf{S},t_0} \hspace{0.3cm} &1/t_0\\
\textnormal{subject to} \hspace{0.3cm} &\frac{1}{2}\left\|\mathbf{S}\right\|_F^2\leq P,\\ 
&t_0\prod_{k=1}^K\left(\frac{g_k(\mathbf{S},\mathbf{\Phi}^{\star})}{\gamma_k}\right)^{-\gamma_k}\leq1,\label{standard_GP_constraint_2}
\end{align}
that can be solved using existing software, e.g.\ CVX \cite{CVX}.

\par It is important to note that the tightness of the upper bound \eqref{UB} heavily depends on the choice of $\left\{\gamma_k\right\}$. Following \cite{Beightler:1976,Chiang:2005}, an iterative procedure can be used where at each iteration the standard GP \eqref{standard_GP}-\eqref{standard_GP_constraint_2} is solved for an updated set of $\left\{\gamma_k\right\}$. Assuming a feasible set of magnitude $\mathbf{S}^{(i)}$ at iteration $i$, compute  $\gamma_k(\mathbf{S}^{(i)})=g_k(\mathbf{S}^{(i)},\mathbf{\Phi}^{\star})/z_{DC}(\mathbf{S}^{(i)},\mathbf{\Phi}^{\star})$ $\forall k$. Then solve problem \eqref{standard_GP}-\eqref{standard_GP_constraint_2} to obtain $\mathbf{S}^{(i+1)}$. Repeat the iterations till convergence.

As noted in \cite{Chiang:2005}, since the original problem is nonconvex, the final solution is not guaranteed to be the global optimum. 

\section{Simulation Results}\label{simulations}
\par We now evaluate the performance gain of the optimized adaptive waveform versus two baselines: a non-adaptive waveform not relying on CSIT and an adaptive waveform relying on CSIT but not requiring the optimization of Section \ref{section_waveform}. A suitable choice of non-adaptive waveform for single antenna WPT was shown in \cite{Trotter:2009,Boaventura:2011} to exhibit high PAPR. It was suggested to choose an in-phase multisine excitation with uniform power allocation. We therefore choose the non-adaptive baseline waveform as $\phi_{n,m}=0$ and $s_{n,m}=1/\sqrt{NM}$ $\forall n,m$. The adaptive baseline waveform is chosen as a matched filter (MF) allocating power to all sinewaves but proportionally to the channel strength, i.e.\ $\phi_{n,m}=-\bar{\psi}_{n,m}$ and $s_{n,m}=c A_{n,m}$ with $c$ a constant to guarantee the power normalization. Hence the difference between the optimized waveform and the one based on MF lies in a different choice of amplitudes. $k_2=0.0034$ and $k_4=0.3829$ have been computed for an operating point $a=0$ and used as such to design the optimized waveform. 
\par We assume a WiFi-like environment at a center frequency of 5.18GHz with a 36dBm EIRP, 2dBi receive antenna gain and 50dB path loss in a large open space/office environment with a NLOS channel power delay profile obtained from model B \cite{Medbo:1998b}. Taps are modeled as i.i.d.\ circularly symmetric complex Gaussian random variables. This leads to an average received power of about -12dBm. The frequency gap is fixed as $\Delta_w=2\pi\Delta_f$ with $\Delta_f=B/N$ with $B=20MHz$ and the $N$ sinewaves are centered around 5.18GHz.
\par
\begin{figure}
\centerline{\includegraphics[width=0.8\columnwidth]{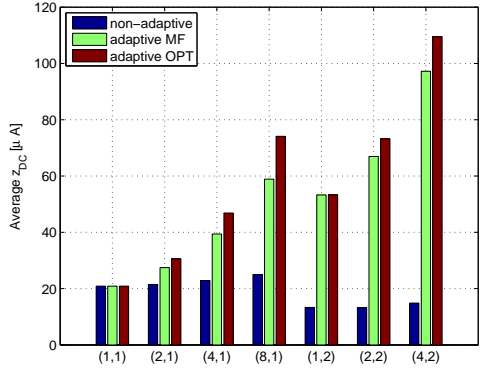}}
  \caption{Average $z_{DC}$ as a function of $(N,M)$.}
  \label{z_DC_results}
\end{figure}

 Fig.\ \ref{z_DC_results} displays the $z_{DC}$ averaged over many channel realizations as a function of $(N,M)$ for the three waveform designs: non-adaptive, adaptive with matched filter MF, adaptive optimized OPT. Significant gains are achieved with adaptive waveforms over non-adaptive ones. The OPT waveform shows an increasing gain over MF as $N$ increases and therefore better exploits the non-linearity of the diode.
\par
\begin{figure}
\centerline{\includegraphics[width=0.8\columnwidth]{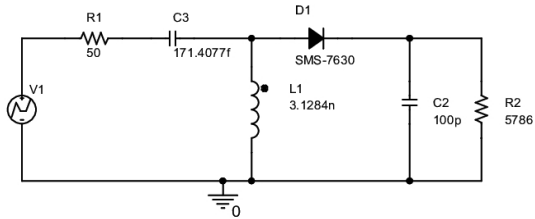}}
  \caption{Rectenna with a single diode and a L-matching network.}
  \label{circuit}
\end{figure}

\begin{figure}
\centerline{\includegraphics[width=0.8\columnwidth]{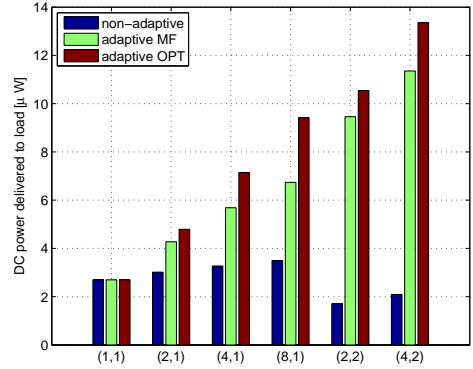}}
  \caption{Average DC power delivered to the load as a function of $(N,M)$.}
  \label{pspice_results}
\end{figure}

 In order to validate the waveform optimization and the EH model, the waveforms have been used as inputs to a realistic energy harvester implemented in PSPICE. In Fig.\ \ref{circuit}, the PSPICE circuit contains a L-matching network \cite{Pinuela:2013} to guarantee a good matching between the rectifier and the antenna. $\textnormal{V1}=v_s(t)=2 y(t) \sqrt{R_{ant}}$ is set as the voltage source. The antenna and load impedances are set as $\textnormal{R1}=R_{ant}=50\Omega$ and $\textnormal{R2}=R_L=5786\Omega$, respectively. The components have been optimized in PSPICE for an input power of -10dBm. Fig.\ \ref{pspice_results} displays the average (over many channel realizations) harvested DC output power. It confirms the observations made in Fig.\ \ref{z_DC_results} and validates the EH model and the waveform optimization. It highlights the significant (and increasing as $N,M$ grow) gains achieved by the optimized waveforms.
\section{Conclusions}\label{conclusions}
The paper derived a methodology to design and optimize multisine waveforms for multi-antenna WPT. Contrary to existing designs, the waveforms are adaptive to the CSI (assumed available to the transmitter), therefore making them more suitable to exploit the non-linearity of the rectifier. They result from a non-convex posynomial maximization problem and are shown through realistic simulations to provide significant gains (in terms of harvested DC power) over state-of-the-art waveforms under a fixed transmit power constraint. The results are expected to trigger significant interests in the RF/WPT community as well as in the communication theory community involved with wireless EH communication and SWIPT.

\ifCLASSOPTIONcaptionsoff
  \newpage
\fi

\end{document}